# Localization of Fetal Head in Ultrasound Images by Multiscale View and Deep Neural Networks

Zahra Sobhaninia, Ali Emami, Nader Karimi, Shadrokh Samavi

*Isfahan University of Technology*
*Isfahan,* 84156-83111 *Iran,*

*Abstract*— One of the routine examinations that are used for prenatal care in many countries is ultrasound imaging. This procedure provides various information about fetus health and development, the progress of the pregnancy and, the baby's due date. Some of the biometric parameters of the fetus, like fetal head circumference (HC), must be measured to check the fetus's health and growth. In this paper, we investigated the effects of using multi-scale inputs in the network. We also propose a light convolutional neural network for automatic HC measurement. Experimental results on an ultrasound dataset of the fetus in different trimesters of pregnancy show that the segmentation accuracy and HC evaluations performed by a light convolutional neural network are comparable to deep convolutional neural networks. The proposed network has fewer parameters and requires less training time.

## I. INTRODUCTION

Medical imaging is one of the robust procedures that can be used for diagnostic and therapeutic purposes. Imaging technologies include magnetic resonance imaging (MRI), Ultrasound (US), medical radiation, and computed tomography (CT) scanners.

Ultrasound imaging is a medical method, which employs high-frequency sound waves to produce dynamic visual images inside the body. Ultrasound imaging operates in real-time and can assist the evaluation, diagnosis, and treatment of diseases in numerous situations. US advantages have made it one of the preferred imaging techniques. Some of these advantages are as follows:

(i) It is safer than other modalities such as X-ray imaging and CT scans.

(ii) It does not require using needles and injections; therefore, it is painless.

(iii) It is widely available and has a lower cost compared to other methods.

(iv) It is used for different purposes, for instance: inspection of heart and blood vessels, breasts, abdomen, muscles, carotid arteries, pregnancy-prenatal diagnosis, and gynecological diseases.

As mentioned above, one of the prenatal cares is using US imaging in different trimesters of pregnancy. It is used to check gestational age calculation, baby's due date, and fetal structures development by measuring biometric parameters, such as baby's abdominal circumference (AC), femur length, humerus length, crown-rump length, biparietal diameter (BPD), and head circumference (HC). HC is measured for estimating its size, weight, and detecting fetus abnormalities [1].

Despite the discussed benefits, ultrasound imaging has its defects, such as the existence of artifacts, attenuation, shadows, speckle noise, missing boundaries, and low signal-to-noise ratio [2]. Some examples of US images that contain noise and incomplete boundaries are shown in Fig 1. These samples show skull area, HC and BPD parameters. The sonographer should find a proper plane; then the caliper must be appropriately positioned. Therefore the procedure efficiency is directly dependent on the operator skill.

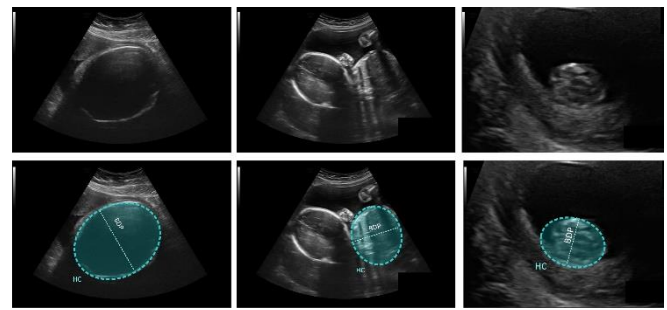

Figure 1 Samples of ultrasound fetal head region (a) Original images (b) HC and BDP provided by a radiologist (blue borders[1]).

To deal with these drawbacks and the time-consuming process, there have been several studies on automated fetal biometrics measurement. Iterative randomized Hough transform (IRHT) is a popular method for detecting ellipse shapes in noisy images without the determination of the skull area or head circumference. However, it requires high computational efforts [3]. Another technique that has been applied for this task is semi-supervised patch-based graphs. It uses non-local information and graph of patches [4].

Some other methods are based on active contour models [5], multilevel thresholding circular shortest paths [6], and morphological operators [7]. There are various machine learning approaches that cope with this task like a probabilistic boosting tree (PBT) applied for AC measuring [8] and random forest classifier using Haar-like features for training and extracting ellipses from segmentations with Hough transform [9].

---

[1] http://doi.org/10.5281/zenodo.1322001

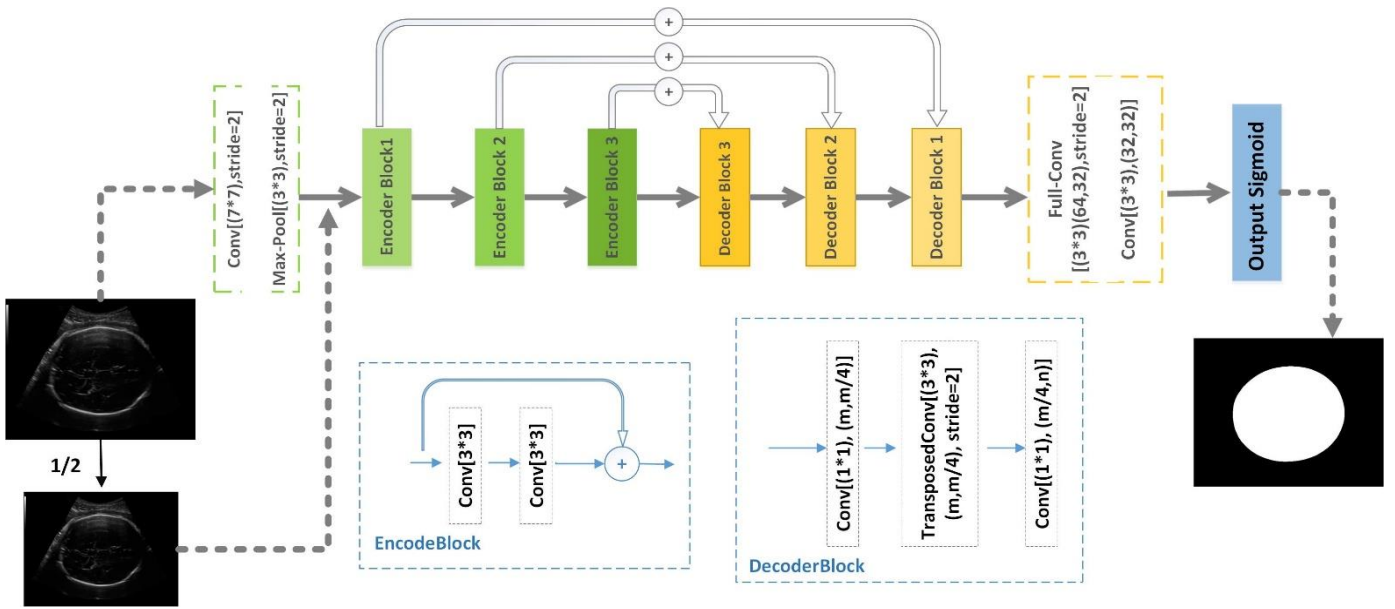

Figure 2  Overview of the proposed network architecture, modified Link-Net called mini-LinkNet

Due to the high performance of convolutional neural networks (CNN) on various image processing tasks [10], recent researches, in the estimation of fetal biometric parameters, has been done with a focus on deep learning approaches. For instance, Jaeseong Jang et al. [11] proposed a CNN structure to analyze images based on anatomical configuration of the umbilical vein and stomach bubble regions in US images. Suitable areas are selected to recognize abdominal area and applied Hough transform is used for the AC measurement [11]. Another study [12] presented a multi-task deep network for fetal biometric parameters estimation. The presented network accomplishes segmentation and ellipse tuner tasks and leads to better performance in comparison with a single task network. It shows training both tasks together helps to improve the accuracy of both tasks [12].

In this paper, we present a CNN based approach to determine the fetal head region in US imaging. The proposed network is a multi-scale and low complexity structure inspired by LinkNet network [13] that has been applied in semantic segmentation. In this work, we show that using a light network can be more helpful for segmentation in some datasets and leads to the desired result. The paper is organized as follows: Section II describes network architecture and then explains the implementation details of fetal head area determination. Section III provides comparative results and discusses the system performance. Finally, we draw the conclusions in Section IV.

## II. PROPOSED METHOD

Figure 2 overviews our network architecture for the fetal head segmentation of US images. In this section, first, we discuss the role of multi-scale inputs in fetal head measurement. Next, we investigate a light LinkNet that improves evaluation criteria. We called this proposed network, mini-LinkNet. After that we discussed the loss function that the network uses.

### A. Proposed Network Structure

The encoder-decoder structure has been preferred to be applied for medical image segmentation because this structure protects detailed information [14]. LinkNet network is an encoder-decoder architecture that is designed for semantic segmentation [13], It also has high performance in medical image segmentation fields [12] [15]. However, there are some shortcomings in these methods such as poor segmentation accuracy in noisy and low contrast images, a high number of trainable network parameters, and need a long time for training them.

As it is shown in Figure 3, LinkNet network comprises encoder and decoder blocks with residual links that connect encoders to decoders. This network uses high-level and low-level feature maps for segmentation. Its structure, in the encoder part, has some downsampling steps that are saved and later used for upsampling in the decoder. In general, spatial information is weakened in this part because of strides in convolutions or because of the pooling mechanism. Therefore in deeper networks, more information on images is lost. Hence it is noteworthy that the network structure is designed in this way that minimizes information loss as much as possible. One of the solutions is considering a half-scale image in addition to the main input image and concatenating it with feature maps (As it is shown in Figure 2). It slightly improves the shortcoming of deep CNN. The results of multi-scale LinkNet improvement are shown in section III.

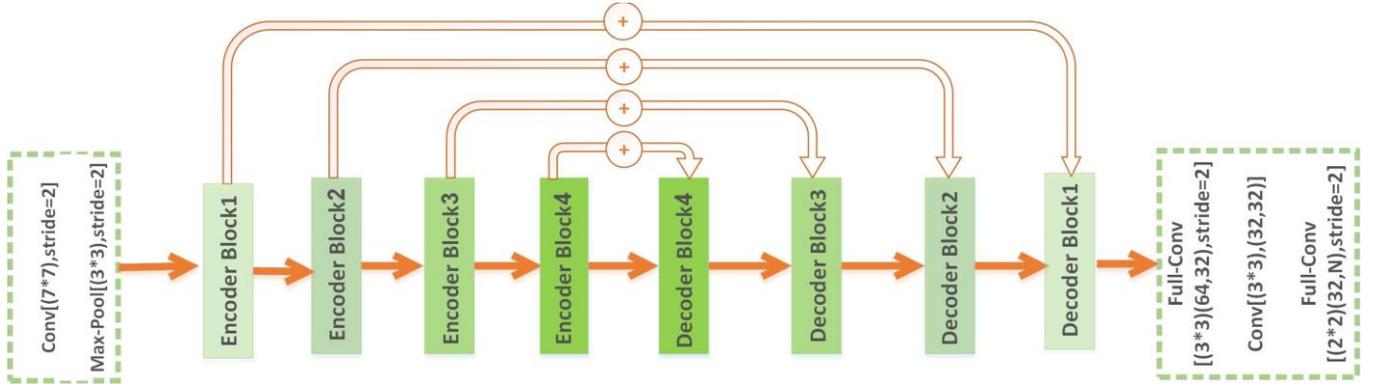

Figure 3  LinkNet Architecture [13]

Another proposed way to overcome the deficiency of deep networks is reducing factors that cause weak information [16]. In this way, we reduce the number of convolutional layers in mini-LinkNet. As it is observed in Fig. 3, there are 4 encoder blocks in the LinkNet network, while mini-LinkNet has only 3 blocks, which seems to be more efficient and could maintain features of the image.

*B. Loss Function*

The common loss function used in medical image segmentation is the Dice [17] which is defined by:

$$Dice = \frac{2\ TP}{2\ TP + FN + FP} \tag{1}$$

where TP denotes true positive, FN is a false negative, and FP represents false positive. However, for challenging US images is better to consider another loss function for network training. We utilize $L_{LN}$, the loss function for a network that is defined by:

$$L_{LN} = w(x) * (Bce) + Dice \tag{2}$$

Where $w$ is weighting representation [18] that defined as:

$$w(x) = 1 + \omega_0 \cdot \exp \frac{d(x)}{2\sigma^2} \tag{3}$$

By considering $\omega(x)$, we enlarge gradient loss on boundaries of fetal head regions. $d(x)$ represents the distance between the ground truth boundaries and pixels. And $\sigma$ is the variance of Gaussian kernel. The amounts of $\omega_0$ and $\sigma$ are considered as 30 and 10, respectively.

$Bce$ is the binary cross-entropy that represents a pixel-based loss metric defined as:

$$Bce(G, I) = -(\ G \times \log(P) + (1 - G) \times \log(1 - P)) \tag{4}$$

Where G is the ground truth, P is the predicted map that is the output of the network.

III. EXPERIMENTAL RESULTS

In this work, we investigate two networks. Both of them were implemented with python and tensor-flow and trained end to end using stochastic gradient descent with momentum (Adam with learning rate = 0.001). The number of trainable parameters of LinkNet network is 11,541,697, while in mini-LinkNet there are 2,894,972 trainable parameters. Hence, the number of trainable parameters is reduced significantly. We ran 150 epochs with considering 10 batch size on NVIDIA GeForce GTX 1080 Ti. LinkNet took 32 hours for training while training time for mini-LinkNet took 18 hours, which shows the training time is almost half by this minimizing.

*A. DataSet*

The dataset that we used contains 999 two-dimensional US images that the size of each of them is 800 by 540 pixels with a pixel size ranging from 0.052 to 0.326 mm. This dataset was collected from the database of the department of obstetrics of the Radboud University Medical Center, Nijmegen, the Netherlands [9].
To increase network efficiency and to prevent overfitting on the training data, we applied data augmentation on the dataset. We used rotation and flip transforms to generate 10 images from each image. Data were randomly divided into an 80% training dataset and 20% validation data.

*B. Evaluation*

We consider difference (DF)(5), the absolute difference (ADF)(6), $Dice$ similarity coefficient, and Hausdorff distance (HD)(7) to assess the performance of our method [9].

$$DF = HC_P - HC_{GT} \tag{5}$$

$$ADF = |HC_P - HC_{GT}| \tag{6}$$

$HC_P$ represents the extracted perimeter from the result of output segmentation and $HC_{GT}$ ground truth fetal head circumference.
HD is the maximum of $h(S, R)$ and $h(R, S)$, which is defined as:

$$HD(S, R) = \max(h(S, R), h(R, S)) \tag{7}$$

where $S = \{s_1, \ldots, s_q\}$ represents pixels from the result of output segmentation and $R = \{r_1, \ldots, r_q\}$ are pixels from the ground truth. $h(S, R)$ is defined [9]:

$$h(S, R) = \max_{s \in S} \min_{r \in R} ||\ s - r|| \tag{8}$$

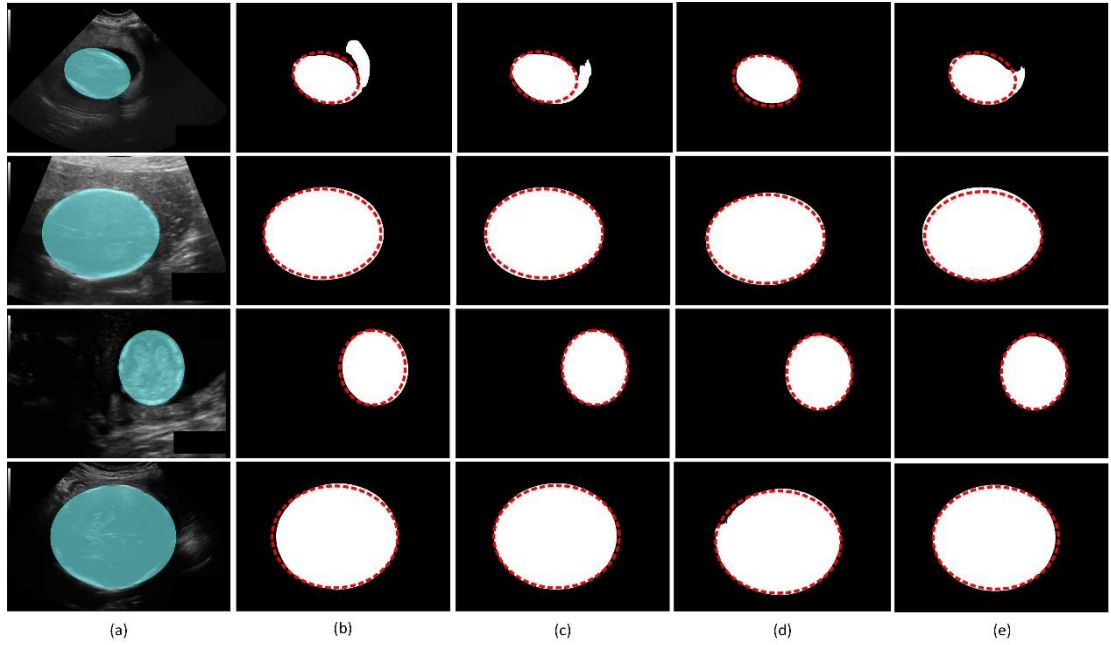

Figure 4 Segmentation results of the fetal head (ground truth is red). From left to right: (a) Original image, (b) results of LinkNet network, (c) MultiScale LinkNet, (d) mini-LinkNet and (e) multi-scale mini-LinkNet network

Table 1 elaborates on the results of applying multiscale inputs to LinkNet on the test dataset. As shown, there are improvements in all evaluation parameters.

TABLE 1 EVALUATION RESULTS OF LINKNET AND MULT-ISCALE LINKNET

| Method | Dice Score % | DF(mm) | ADF(mm) | HD (mm) |
|---|---|---|---|---|
| LinkNet | 91.60 | 1.92 | 4.95 | 4.81 |
| MultiScale LinkNet | **93.75** | **1.53** | **2.27** | **3.70** |

Table 2 demonstrates the mini-LinkNet evaluation parameters and the effects of lighting LinkNet. Although mini-LinkNet has lower layers in comparison with LinkNet, it has a better performance in less time on US imaging as shown in the second row in Table 2 that four parameters improved. Furthermore, it investigates the role of multi-scaling on mini-LinkNet, and the third row shows the Improvement of ADF and HD parameters.

TABLE 2 COMPARISION OF LINKNET AND MINI-LINKNET

| Method | Dise Score % | DF(mm) | ADF(mm) | HD (mm) |
|---|---|---|---|---|
| LinkNet | 91.60 | 1.92 | 4.95 | 4.81 |
| mini-LinkNet | **92.65** | **0.94** | 2.39 | 3.53 |
| MultiScale mini-LinkNet | 92.46 | 1.19 | **2.22** | **3.40** |

Fig. 4 shows segmentation results with different networks. It shows the improvement of segmentation by adding scale to Linknet. A comparison of the (b) and (c) columns shows this point. Also, it elaborates minimizing the layers of the network improves all evaluation parameters of the segmentation. A comparison of (b) and (d) columns elaborates this issue. The effect of adding multi-scale to mini-LinkNet Can be observed by comparing the (d) and (e) columns.

Conclusion

This paper discussed convolutional neural network methods for automatic fetal head segmentation. At first, it investigates multi-scale effects on LinkNet performance of US imaging, then it presented a mini-LinkNet network for this task. Evaluation criteria are improved with this approach in comparison with deep LinkNet, while the number of trainable parameters and training time is lower.

References

[1] Pam Loughna , Lyn Chitty , Tony Evans , Trish Chudleigh, "Fetal Size and Dating: Charts Recommended for Clinical Obstetric Practice," Ultrasound, vol. 17, no. 3, pp. 160-166, 2009.

[2] Sylvia Rueda, Sana Fathima, Caroline L. Knight, Mohammad Yaqub Aris T. Papageorghiou, "Evaluation and Comparison of Current Fetal Ultrasound Image Segmentation Methods for Biometric Measurements: A Grand Challenge," IEEE Transactions on Medical Imaging, pp. 797 - 813, April 2014.

[3] WeiLu, JingluTan, Randall Floyd, "Automated fetal head detection and measurement in ultrasound images by iterative randomized hough transform," Ultrasound in Medicine & Biology, vol. 31, no. 7, pp. 929-936, 2005.

[4] Ciurte A, Bresson X, Cuadra MB, "A semi-supervised patch-based approach for segmentation of fetal ultrasound imaging," in Proceedings of Challenge US: Biometric Measurements from Fetal Ultrasound Images ISBI, 2012.

[5] J. L. Perez-GonzalezEmail, C. Bello MuńozM, C. Rolon Porras Fernando et al., "Automatic Fetal Head Measurements from Ultrasound Images Using Optimal Ellipse Detection and Texture Maps," Springer, vol. 49, p. 329–332, 2015.

[6] Ponomarev GV, Gelfand MS, Kazanov MD, "A multilevel thresholding combined with edge detection and shape-based recognition for segmentation of fetal ultrasound images," in Proceedings of Challenge US: Biometric Measurements from Fetal Ultrasound Images, ISB, 2012.

[7] Vibhakar Shrimali, R. S. Anand, Vinod Kumar, "Improved segmentation of ultrasound images for fetal biometry, using morphological operators," in Annual International Conference of the IEEE Engineering in Medicine and Biology Society, Sept. 2009.

[8] Gustavo Carneiro, Bogdan Georgescu ,Sara Good ,Dorin Comaniciu, "Detection and measurement of fetal anatomies from ultrasound images using a constrained probabilistic boosting tree," IEEE

Transactions on Medical Imaging (, vol. 27, no. 9, pp. 1342 - 1355, Sept. 2008.

[9] Thomas L. A. van den Heuvel, Dagmar de Bruijn, et al., "Automated measurement of fetal head circumference using 2D ultrasound images," PLoS ONE, 2018.

[10] Geert Litjens, Thijs Kooi, Babak Ehteshami Bejnordi, et al., "A survey on deep learning in medical image analysis," Medical Image Analysis, vol. 42, pp. 60-88, 2017.

[11] Jaeseong Jang ,Yejin Park ,Bukweon Kim, Sung Min Lee ; Ja-Young Kwon, Jin Keun Seo, "Automatic Estimation of Fetal Abdominal Circumference From Ultrasound Images," IEEE Journal of Biomedical and Health Informatics, vol. 22, no. 5, pp. 1512 - 1520, 2017.

[12] Zahra Sobhaninia, Shima Rafiei , Ali Emami , Nader Karimi ,Kayvan Najarian , Shadrokh Samavi, "Fetal Ultrasound Image Segmentation for Measuring Biometric Parameters Using Multi-Task Deep Learning," in 41st Annual International Conference of the IEEE Engineering in Medicine and Biology Society (EMBC), Berlin, Germany, 2019.

[13] Abhishek Chaurasia, Eugenio Culurciello, "LinkNet: Exploiting encoder representations for efficient semantic segmentation," in IEEE Visual Communications and Image Processing (VCIP), St. Petersburg, FL, USA, 2017.

[14] Olaf Ronneberger,Philipp Fischer,Thomas Brox, "U-Net: Convolutional Networks for Biomedical Image Segmentation," International Conference on Medical Image Computing and Computer-Assisted Intervention, pp. 234-241, 2015.

[15] Zahra Sobhaninia, Safiyeh Rezaei, Alireza Noroozi, Mehdi Ahmadi, et al., "Brain tumor segmentation using deep learning by type specific sorting of images," 2018.

[16] Yun Wang,Chang Wei, Zhi-Jian Wang, Qing-Gao Lu, Cheng-Gang Wang, "A More Streamlined U-net for Nerve Segmentation in Ultrasound Images," in Chinese Automation Congress (CAC), 2 Dec. 2018.

[17] Carole H. Sudre, Wenqi Li,Tom Vercauteren,Sebastien Ourselin,M. Jorge Cardoso, "Generalised Dice Overlap as a Deep Learning Loss Function for Highly Unbalanced Segmentations," in International Workshop on Deep Learning in Medical Image Analysis, 2017.

[18] Shima Rafiei, Ebrahim Nasr-Esfahani, Kayvan Najarian, et al., "Liver Segmentation in CT Images Using Three Dimensional to Two Dimensional Fully Convolutional Network," in IEEE International Conference on Image Processing (ICIP), 2018.

[19] Juan J. Cerrolaza, Matthew Sinclair, Yuanwei Li, et al., "Deep Learning with Ultrasound Physics for Fetal Skull Segmentation," in 15th International Symposium on Biomedical Imaging (ISBI), Washington, D.C., USA, 2018.

[20] Matthew D.Sinclair, Juan CerrolazaMartinez, Emily Skelton, et al., "Cascaded Transforming Multi-task Networks For Abdominal Biometric Estimation from Ultrasound," in 1st Conference on Medical Imaging with Deep Learning (MIDL ), Amsterdam, The Netherlands, 2018.